\documentclass[preprint,12pt]{elsarticle}

\usepackage{outlines}
\usepackage{amsmath}
\usepackage{amssymb}
\usepackage{physics}
\usepackage{array}
\usepackage[final]{changes}
\def\hc{\text{h.c.}}
\def\dim{\text{dim}}
\def\S{\text{S}}
\def\O{\text{O}}
\def\SO{\text{SO}}
\def\C{\text{C}}

\def\r{\mathbf r}
\def\rp{\mathbf{r'}}

\usepackage{amssymb}

\journal{Computer Physics Communications}

\graphicspath{{Irudiak/}}

\begin{document}

\begin{frontmatter}

\title{The PointGroupNRG code for
numerical renormalization group calculations
with discrete point-group symmetries.}
\author[EHU,DIPC]{Aitor Calvo-Fernández}
\author[PMA,DIPC,CFM]{María Blanco-Rey}
\author[EHU,DIPC,QC]{Asier Eiguren}
\address[EHU]{
    Departamento de Física, 
    Universidad del País Vasco UPV-EHU, 
    48080 Leioa,
    Spain
}
\address[DIPC]{
    Donostia International Physics Center (DIPC), 
    20018 Donostia-San Sebastián, 
    Spain
}
\address[PMA]{
    Departamento de Polímeros y Materiales Avanzados: Física, Química y Tecnología, 
    Universidad del País Vasco UPV-EHU, 
    20018 Donostia-San Sebastián, 
    Spain
}
\address[CFM]{
    Centro de Física de Materiales CFM/MPC (CSIC-UPV/EHU), 
    Paseo Manuel de Lardizábal 5, 
    20018 Donostia-San Sebastián,
    Spain
}
\address[QC]{
    EHU Quantum Center, 
    University of the Basque Country UPV/EHU,
    48940 Leioa, 
    Spain
}

\begin{abstract}
The numerical renormalization group (NRG) has been widely
used as a magnetic impurity solver since the pioneering
works by Wilson. Over the past decades, a significant
attention has been focused on the application of symmetries
in order to reduce the computational cost of the
calculations and to improve their accuracy. In particular, a
notable progress has been made in implementing continuous
symmetries such as $SO(3)$, useful for studying impurities in
an isotropic medium, or $SU(N)$, which is applicable to a
wide range of systems. In this work, we focus on the
application of discrete point group symmetries, which are
particularly relevant for impurity systems in metals where
crystal field effects are important. With this aim, we have
developed an original NRG code written in the Julia
language, \verb|PointGroupNRG|, where we have implemented
crystal point-group symmetries for the Anderson impurity
model, as well as the continuous spin and charge
symmetries. Among other results, we demonstrate the advantage
of our procedure by \added{applying the code to
a} \deleted{performing thermodynamic calculations for an}
\added{two-impurity model with RKKY interaction and an}
impurity system with two orbitals of $E_g$ symmetry and two
channels. \deleted{We compare the results with those obtained for an
approximate equivalent model with continuous orbital
symmetry.} \added{We also provide benchmarks to show the
performance improvements obtained by exploiting the orbital symmetries.}

\end{abstract}


\end{frontmatter}


\section{Introduction}

\added{
The Numerical Renormalization Group (NRG) is a
non-perturbative method pioneered by K. G. Wilson to solve
the single-channel spin-$1/2$ Kondo model, which consists of
a spin degree of freedom coupled to a bath of conduction
electrons \cite{wilson1975}. The method was devised in order
to overcome the limitations of the perturbative treatment of
the Kondo model, which results in infrared divergences
for various physical quantities \cite{Hewson}. The strategy
followed by the NRG is to perform a logarithmic
discretization of the continuum of conduction
degrees of freedom, dividing it into intervals 
$[\Lambda^{-(n+1)},\Lambda^{-n}]$ and
$[-\Lambda^{-n},-\Lambda^{-(n+1)}]$ for $n=0,1,2,\dots$
for a discretization parameter
$\Lambda>1$, where the coupling terms introduced by each of 
these intervals decrease exponentially with $n$. The
NRG method proceeds from the highest-energy excitations to
the lowest energy scales by iteratively adding these
intervals to the Hamiltonian, diagonalizing it, and
discarding the highest-energy states.
}
\par
\added{
Since its first implementation, the NRG method has seen
improvements and extensions in various aspects. Some
limitations of the original discretization scheme have been
overcome by utilizing, for instance, a
non-orthogonal conduction basis that
removes the artificial renormalization of the impurity-band
hybridization \cite{campo2005} or an
adaptive discretization mesh that allows for a more accurate
modeling of conduction bands with
a non-constant density of states \cite{zitko2009b}. 
Other methodological developments have been aimed at
aleviating the computational cost of the calculations, which
becomes an obstacle as the number of impurities or the
degrees of freedom of the impurity and the bath increase.
One improvement in this respect is the combination of
calculations performed for various interleaved
discretizations, which allows to use large values
of the discretization parameter $\Lambda$ and, at the same
time, obtain smooth spectral functions \cite{yoshida1990} and
thermodynamic functions \cite{oliveira1994} 
}
\par 
\deleted{
Symmetries have been implemented in the Numerical
Renormalization Group method since the earliest works. In
the seminal paper by Krishna-Murthy et al.
\cite{krishna-murthy1980}, $SU(2)$ spin-symmetry was used in
order to reduce the computational cost of the calculations,
which were performed for a single-orbital, single-channel
Anderson model. Since then, the attention has been directed
to ever more complex magnetic impurity models and systems
with multiple orbitals and/or channels. In those cases,
symmetry is often used to improve the efficiency of the
calculations. Notable developments in this area include the
explicit application of continuous non-Abelian symmetries:
(i) $SU(N)$ (special unitary) for the orbital, spin-orbital
and charge sectors
\citep{moca2012,filippone2014,mantelli2016}; (ii) $Sp(N)$
(symplectic) for the spin-charge sector
\cite{weichselbaum2012}, and (iii) $SO(3)$ (rotational) for
the orbital sector}
\par
\added{
Another line of improvement, which we follow in this work,
is the application of symmetries. Symmetries have been
implemented in the NRG method since the earliest works. In
the seminal paper by Krishna-Murthy et al.
\cite{krishna-murthy1980}, charge and spin symmetries were
used in order to reduce the computational cost of the
calculations, which were performed for a single-orbital,
single-channel Anderson model. The extension of the NRG to
more complex models, often featuring more degrees of
freedom, has motivated the implementation of various
continuous non-Abelian symmetries: (i) $SU(N)$ (special
unitary) for the orbital, spin-orbital and charge sectors
\citep{moca2012,filippone2014,mantelli2016}; (ii) $Sp(N)$
(symplectic) for the spin-charge sector
\cite{weichselbaum2012}, and (iii) $SO(3)$ (rotational) for
the orbital sector \cite{sakai1989,zitko2009}.  
}
\par 
In this paper we present \verb|PointGroupNRG|
 \cite{pointgroupnrg}, a
Julia \cite{julialang} code that implements finite point group
symmetries in Anderson Hamiltonians. In particular, our code
is designed to work with symmetries of the form $G =
U(1)_\C \otimes P_\O \otimes SU(2)_\S$, where $U(1)_\C$ is
the charge symmetry corresponding to particle conservation,
$P_\O$ is a simply reducible finite orbital
point-group symmetry, and $SU(2)_\S$ represents spin
isotropy. These symmetries are appropriate for systems where
spin-orbit interactions can be neglected and thus the
orbital and spin sectors can be treated separately. Simply
reducible point groups include common symmetries such as the
cubic crystal point groups and the inversion symmetry. More 
specifically, the main features of the code are (i) the
automatic construction of symmetry-adapted multiplets, (ii)
tools to generate fully symmetric Anderson impurity
Hamiltonians, and (iii) an iterative diagonalization
procedure that exploits the specified symmetries to increase
the efficiency and precision of the calculations.
\added{The PointGroupNRG package is open source. It can be
downloaded from \cite{pointgroupnrg} alongside a manual, a
tutorial, and scripts to build an optional precompiled
version. In addition to the singular symmetry-handling
features, it also includes modern discretization techniques.}
\par 
The paper is organized as follows. The first two sections
are devoted to those aspects of the procedure that are most
specific to our code and the type of symmetries it deals
with: Section \ref{multiplets} covers the automatic
generation of symmetry-adapted multiplet states and Section
\ref{Hamiltonian} deals with the generation of point-group
symmetric Hamiltonians. In Section \ref{implementation} we
briefly cover the implementation of the NRG procedure, which
follows the approach described in the existing published
literature. \deleted{We finish in Section \ref{Eg} by showing results
for a two-orbital $E_g$ model, which highlights the
advantages of incorporating discrete point-group
symmetries.} \added{In Section \ref{Applications} we
showcase the application of \texttt{PointGroupNRG} with 
finite orbital symmetries to a
two-impurity system exhibiting RKKY interaction (Section 
\ref{RKKY}) and a two-orbital $E_g$ model (Section
\ref{Eg}). Finally, Section \ref{Benchmarks} contains
benchmarks that demonstrate the computational advantages of 
explointing orbital symmetries with \texttt{PointGroupNRG} 
(Section \ref{benchmarks_symmetry}) and a performance 
comparison of \texttt{PointGroupNRG} with another code
(Section \ref{benchmarks_comparison}).} 

\section{Symmetry-adapted basis} \label{multiplets}
Our magnetic impurity system has symmetry $G =
U(1)_\C \otimes P_\O \otimes SU(2)_\S$. To fully exploit
it, we choose to work with a symmetry-adapted basis from the
outset, so we define symmetry-adapted states 
\begin{equation}
    \ket{w} = \ket{\Gamma_w,\gamma_w,r_w},
\end{equation}
where the quantum numbers $\Gamma_w$, $\gamma_w$ and $r_w$
label the irrep, the partner and the outer multiplicity,
respectively. Our notation is based on Ref. \cite{moca2012}.
Due to the tensor-product structure of the group $
G$, the irreps can be decomposed as
$\Gamma_w=(N_w,I_w,S_w)$, where $N_w$ is the particle
number, $I_w$ is the orbital irrep, and $S_w$ is the total
spin. We label the partners accordingly as
$\gamma_w=(i_w,s_w)$, where $i_w$ is the orbital partner
label and $s_w$ is the spin projection (we do not explicitly
label charge partners because the irreps $N_w$ of $U(1)_\C$ are
one-dimensional). 
\par
The building blocks for constructing the initial basis and
updating it at each NRG step are the multiplet states for
the impurity and the shells, \textit{i.e.}, the
conduction-channel sites \cite{krishna-murthy1980}. Our
\verb|PointGroupNRG| code generates these multiplet states
for the impurity and shell subspaces by taking as the only
input the relevant Clebsch-Gordan coefficients of the point
symmetry group. Here we briefly outline the main steps in
the procedure (see also Refs. \cite{Hamermesh,Tung,Dresselhaus}). 
\par
We first construct $N$-particle orbital and spin wave
functions. Orbital (spin) basis functions have the form 
$\ket{\psi}_N=\ket{I_\psi,i_\psi,r_\psi}_N$ 
($\ket{\xi}_N=\ket{S_\xi,s_\xi,r_\xi}_N$), where $I_\psi$
and $i_\psi$ ($S_\xi$ and $s_\xi$) are orbital (spin) irrep
and partner quantum numbers, respectively, and $r_\psi$
($r_\xi$) is the outer multiplicity. The construction is
carried out recursively by means of the Clebsch-Gordan
series
\begin{equation}
\begin{aligned}
    &\ket{\psi}_N = 
    \sum_{i_{\psi'},i_{\psi''}} 
    (I_{\psi'},i_{\psi'};I_{\psi''},i_{\psi''}|I_\psi,i_\psi,r_\psi)
    \ket{\psi'}_{N-1}\otimes
    \ket{\psi''}_1,
    \\
    &\ket{\xi}_N = 
    \sum_{s_{\xi'},s_{\xi''}} 
    (S_{\xi'},s_{\xi'};S_{\xi''},s_{\xi''}|S_\xi,s_\xi,r_\xi)
    \ket{\xi'}_{N-1}\otimes\ket{\xi''}_1,
\end{aligned}
\end{equation}
where the $(\dots;\dots|\dots)$ terms are Clebsch-Gordan
coefficients.
\par
In order to find the antisymmetric combinations, we
implement the apparatus of Young theory \cite{Tung}. Since
the $N$-element permutation group $\mathcal S_N$ commutes with the
orbital (spin) symmetry group $P_\O$ ($SU(2)_\S$), we can further
subdivide each subspace of states with quantum
numbers $I_\psi$ and $i_\psi$ ($S_\xi$ and $s_\xi$) by introducing permutation
irrep and partner quantum numbers, $A$ and $a$ ($B$ and $b$). We do
so by going through all the orbital (spin) states
and applying the Young symmetrizer $Y_{A,a}$ ($Y_{B,b}$),
which is an operator. The result can be either zero or a
wave function of the form 
\begin{equation}
\begin{aligned}
    &Y_{A,a} \ket{I_\psi,i_\psi,r_\psi}_N =
    \ket{I_\psi,i_\psi,r_{\psi'};A,a} 
    :=\ket{\psi}_N^{A,a}
    \\
    &Y_{B,b} \ket{S_\xi,s_\xi,r_\xi}  =
        \ket{S_\xi,s_\xi,r_{\xi'};B,b}
    :=\ket{\xi}_N^{B,b},
\end{aligned}
\end{equation}
where the state $\ket{\psi}_N^{A,a}$ ($\ket{\xi}_N^{B,b}$)
transforms as the $a$ ($b$) partner of the $A$ ($B$) irrep
of $\mathcal S_N$. A set of linearly independent states
$\ket{\psi}_N^{A,a}$ ($\ket{\xi}^{B,b}_N$) provides a
complete basis of states adapted to the orbital (spin) and
permutation symmetries. 
\par
Antisymmetric spin-orbital states result from
combining orbital and spin states belonging to
permutation irreps $A$ and $B$ such that the antisymmetric
permutation irrep $\bar C$ is contained in the product 
$C=A\otimes B$,
\begin{equation}
    C = A \otimes B = \dots + \bar n \bar C + \dots,
\end{equation}
where $\bar n$ is the multiplicity of $\bar C$,
\textit{i.e.}, the number of
times $\bar C$ appears in the product $C$. Therefore, in order
to find the antisymmetric wave functions we do as follows.
For each group of quantum numbers
$Q=(N,I_\psi,i_\psi,S_\xi,s_\xi,A,B)$ we
construct a basis composed of all tensor product wave functions
$\ket{I_\psi,i_\psi,r_\psi}_N^{A,a} \otimes 
\ket{S_\xi,s_\xi,r_\xi}_N^{B,b}$ with varying quantum 
numbers $a$, $b$, $r_\psi$ and $r_\xi$. This basis spans a
space $V_C^{(Q)}$ of the representation $C$ that conserves
the quantum numbers $Q$. In order to find the subspace
belonging to irrep $\bar C$, $V_{\bar C}^{(Q)}\subseteq V_C^{(Q)}$, 
we first construct a basis of the space $V_{\bar C}$ of all 
antisymmetric spin-orbital states. The antisymmetric
subspace with quantum numbers $Q$ is the intersection 
$V_{\bar C}^{(Q)}=V_C^{(Q)}\cap V_{\bar C}$, which we compute
numerically to obtain the basis functions
\begin{equation}
\begin{aligned}
    \ket{w} 
    = 
    \ket{\Gamma_w,\gamma_w,r_w}
    =
    \sum_{r_\psi r_\xi}
    c_Q(r_\psi,r_\xi;w)
    \ket{I_w,i_w,r_\psi}^{A,a}_{N_w} 
    \otimes 
    \ket{S_w,s_w,r_\xi}^{B,c}_{N_w},
\end{aligned}
\end{equation}
where the antisymmetric state $\ket{w}$ is expressed in the
spin-orbital tensor product basis of the $V_C^{(Q)}$
subspace, being $c_Q(r_\psi,r_\xi;w)$ the coefficients. With
a change of basis, we re-express $\ket{w}$ in the basis of
Fock states as
\begin{equation}
\begin{aligned}
    \label{eq:fock_coefficients}
    \ket{w}= 
    \sum_{\mathfrak a} 
    c_F(\mathfrak a;w)
    \big(\prod_{\alpha_i\in\mathfrak a}
    f^\dagger_{\alpha_i}\big)
    \ket{\Omega},
\end{aligned}
\end{equation}
where $c_F(\mathfrak a;w)$ is the coefficient of the $N_w$ particle
Slater determinant state with occupations $\mathfrak a =
(\alpha_1,\dots,\alpha_{N_w})$, being $\alpha_i$ the label
for one-particle symmetry-adapted states; $f^\dagger_{\alpha_i}$ creates a
particle in state $\alpha_i$, and $\ket{\Omega}$
represents an empty impurity or shell. Multiplet Fock
states $\ket{w}$ for an impurity and a shell, or for
different shells, can then be combined
by a regular Clebsch-Gordan series (\textit{i.e.}, regardless
of permutation symmetries)
in order to iteratively form states along the NRG sequence
(see Section \ref{App.A}).

\section{Construction of a symmetric Hamiltonian} \label{Hamiltonian} 
Let us consider a multi-orbital, multi-channel Anderson
Hamiltonian
\begin{equation}
    \begin{aligned}
        & \mathcal H 
        = H_\text{occ} + H_\text{C} + H_\text{hyb} + H_\text{cond}, \\ 
        & H_\text{occ}
        = \sum_{\alpha} \epsilon_\alpha
            f^\dagger_{\alpha} 
            f_{\alpha}, \\
        & H_\C
        = \sum_{\alpha_1\alpha_2\alpha_3\alpha_4}
           U_{\alpha_1\alpha_2\alpha_3\alpha_4} 
           f^\dagger_{\alpha_1} f^\dagger_{\alpha_2}
           f_{\alpha_3} f_{\alpha_4}, \\
        & H_\text{hyb} 
        = \sum_{\alpha\beta} \int_{-D}^D 
        V_{\alpha\beta}[\rho_\beta(\epsilon)]^\frac{1}{2}
            ( f_{\alpha}^\dagger c_{\epsilon\beta}
            +\text{h.c.} ) d\epsilon, \\
        & H_\text{cond}
        = \sum_{\beta} \int_{-D}^D 
            c^\dagger_{\epsilon\beta} 
            c_{\epsilon \beta } 
            \epsilon d\epsilon.
    \end{aligned}
\end{equation}
In the definitions above, the operators $f_{\alpha}$
annihilate impurity electrons in the one-particle
symmetry-adapted impurity state $\ket{\alpha}$; operators
$c_{\epsilon\beta}$ annihilate conduction electrons with
one-particle symmetry quantum numbers $\beta$ and energy
$\epsilon\in[-D,D]$, where $D$ is the half-bandwidth. The
Hamiltonian is broken down into the following terms: $H_\text{occ}$ is the
occupation term, regulated by the occupation energies
$\epsilon_\alpha$; $H_\C$ is the screened Coulomb repulsion
between electrons, given by the parameters
$U_{\alpha_1\alpha_2\alpha_3\alpha_4}$; $H_\text{hyb}$ is the
hybridization term, given by the hybridization
amplitudes $V_{\alpha\beta}$ and 
\replaced{
the density of states (DOS) $\rho_\beta(\epsilon)$ for each
channel $\beta$
}{
a constant density of states $\rho=\frac{1}{2D}$ 
}, and $H_\text{cond}$
is the conduction term.
\par 
We impose symmetry restrictions on the \deleted{parameters
of the}
Hamiltonian by applying the selection rule 
\begin{equation}
    \label{eq:selection_rule}
    \bra{w}\mathcal H\ket{w'}
    \propto 
    \delta_{\Gamma_w,\Gamma_{w'}}
    \delta_{\gamma_w,\gamma_{w'}},
\end{equation}
which makes it explicit that $\mathcal H$ is diagonal in the
irrep and partner quantum numbers. This immediately imposes
the restrictions for the relevant parameters
\added{and DOS functions} in the one-body terms
$H_\text{occ}$ and $H_\text{hyb}$,
\begin{equation}
\begin{aligned}
    &\epsilon_\alpha = \epsilon_{r_\alpha}(\Gamma_\alpha),
    \\
    &V_{\alpha\beta}
    = 
    \delta_{\Gamma_\alpha,\Gamma_\beta}
    \delta_{\gamma_\alpha\gamma_\beta}V_{r_\alpha
    r_\beta}(\Gamma_\alpha),
    \\
    &\added{\rho_\beta(\epsilon) =
    \rho_{r_\beta}(\Gamma_\beta;\epsilon),}
\end{aligned}
\label{eq:epsilon_V_rho}
\end{equation}
where the dependence on $\Gamma_\alpha$ and $\Gamma_\beta$
comes only from the orbital part, because
$N_\alpha=N_\beta=1$ and $S_\alpha=S_\beta=\frac{1}{2}$. The
independent one-body parameters \added{and
functions} are thus
$\epsilon_{r_\alpha}(\Gamma_\alpha)$\replaced{,}{ and}
$V_{r_\alpha r_\beta}(\Gamma_\alpha)$
\added{and }
$\added{\rho_{r_\beta}(\Gamma_\beta;\epsilon)}$.

\par 
For the two-body Coulomb
interaction $H_\C$, the selection rule given by Eq.
\ref{eq:selection_rule} must be applied to two-electron
symmetry-adapted states 
\begin{equation}
    \ket{\kappa}
    = 
    \sum_{\alpha_1\alpha_2} 
    c(\alpha_1,\alpha_2;\kappa)
    f^\dagger_{\alpha_1} 
    f^\dagger_{\alpha_2}
    \ket{\Omega},
\end{equation}
where $c(\alpha_1,\alpha_2;\kappa)$ are the coefficients in
Eq. \ref{eq:fock_coefficients}. Overall, we have that
\begin{align}
    U_{\alpha_1\alpha_2\alpha_3\alpha_4} 
    &=
    \bra{\Omega}f_{\alpha_2}f_{\alpha_1}
    H_\C 
    f^\dagger_{\alpha_4}f^\dagger_{\alpha_3}\ket{\Omega}
    \nonumber \\
    &=
    \sum_{\kappa,\kappa'} 
    c^*(\alpha_1,\alpha_2;\kappa)
    c(\alpha_4,\alpha_3;\kappa')
    \bra{\kappa}H_\C \ket{\kappa'}
    \nonumber \\
    &=
    \sum_{\kappa,\kappa'} 
    c^*(\alpha_1,\alpha_2;\kappa)
    c(\alpha_4,\alpha_3;\kappa')
    \delta_{\Gamma_\kappa \Gamma_{\kappa'}}
    \delta_{\gamma_\kappa \gamma_{\kappa'}}
    U_{r_\kappa r_{\kappa'}}(\Gamma_\kappa),
    \label{Utransformation}
\end{align}
where $U(\Gamma_\kappa)$ is a real symmetric matrix of real
parameters, $U_{r_\kappa r_{\kappa'}} (\Gamma_\kappa) =
U_{r_{\kappa'} r_\kappa} (\Gamma_\kappa)$. Furthermore,
since the orbital basis can be chosen to be real, additional
restrictions apply on top of the restrictions derived from
the symmetry group $G$. In the basis of real orbitals
$\psi_d(\r)$, the Coulomb parameters are
\begin{equation}
    U_{d_1 d_2 d_3 d_4} 
    = 
    \int d\mathbf r d\mathbf r' 
    \psi_{d_1}(\mathbf r) \psi_{d_2}(\mathbf r') 
    H_{\C} (\mathbf r,\mathbf r')
    \psi_{d_3}(\mathbf r') \psi_{d_4}(\mathbf r).
    \label{eq:CoulombIntegral}
\end{equation}
The parameters $U_{d_1 d_2 d_3 d_4}$ have to reflect the
symmetry under the group of permutations in the $d_i$
indices leaving the integral in Eq. \ref{eq:CoulombIntegral}
invariant \deleted{(see Ref. \cite{bunemann2017})}. The restrictions
imposed by this permutation symmetry have to be carried over
to the symmetry-adapted parameters $U_{r_\kappa
r_{\kappa'}}(\Gamma_\kappa)$; this step depends on the
orbital basis chosen and on the form of the multiplet states
generated by the code.
\par
Importantly, this procedure for constructing $H_\C$ only
imposes the finite point-group symmetry to the Coulomb
interaction, so it allows us to construct general
non-spherical screened interaction terms $H_C$ with
$H_\C(\r,\rp)\neq H_\C(|\r-\rp|)$. For a more comprehensive
discussion about Coulomb terms with finite point group
symmetries, see Ref. \cite{bunemann2017}.
\par 
In summary, we are left with a set of parameters \added{and
DOS functions} $\epsilon_{r_\alpha}(\Gamma_\alpha)$,
$V_{r_\alpha r_\beta}(\Gamma_\alpha)$\added{,}\deleted{and}
$U_{r_\kappa r_\kappa'}(\Gamma_\kappa)$ and
$\rho_{r_\beta}(\Gamma_\beta;\epsilon)$ that are model
specific and can be freely chosen; the implemented
construction procedure ensures that the resulting
Hamiltonian is hermitic and respects the
\replaced{symmetries}{permutation symmetry} of the Coulomb
interaction. Our \verb|PointGroupNRG| code takes these
parameters as input in order to construct an Anderson
Hamiltonian with the desired symmetry.

\section{Implementation of the NRG} \label{implementation}
The first step in the NRG procedure is to discretize the
conduction band. We use an interleaved discretization
\cite{yoshida1990,campo2005}, which depends on the
discretization parameter $\Lambda$, commonly used in NRG
calculations \cite{krishna-murthy1980}, and the so-called
shift parameter $z\deleted{\in[0,1)}$ . \deleted{This allows us to average over
independent calculations for different values of $z$,
thereby improving the numerical accuracy of the
thermodynamic curves and spectral functions.} \replaced{Using a Lanczos
algorithm modified for improved numerical robustness
\cite{chen1998}, the continuous conduction term
$H_\text{cond}$ is discretized as}{The discretization transforms the
conduction part $H_\text{cond}$ as}
\begin{equation}
    \begin{aligned}
        \replaced{
            \int c^\dagger_{\epsilon \beta}
            c_{\epsilon \beta} \epsilon
            d\epsilon
        }{
            \int c^\dagger_{\epsilon \beta}
            c_{\epsilon m \sigma} \epsilon
            d\epsilon
        }
        \rightarrow
        \replaced{
           \sum_\beta
            \sum_{n=0}^\infty
            \left[
                h_{n\beta}^{(z)}(c^\dagger_{n\beta}c_{n+1,\beta} 
                + \text{h.c.})
                + e_{n\beta}^{(z)} c^\dagger_{n\beta}c_{n\beta}
            \right]
        }{
            \sum_{n=1}^\infty h_n^{(z)}(c^\dagger_{\beta,n}c_{\beta,n-1} 
            + \text{h.c.})
        },
    \end{aligned} 
\end{equation}
where \added{$c^\dagger_{n\beta}$ ($c_{n\beta}$) creates
(annihilates) an electron at the $n$-th conduction shell,}
$h_{n\beta}^{(z)}$ are the hopping factors \deleted{for a simple
band with a linear dispersion relation $\epsilon_{\mathbf
k}=Dk$}\added{, and $e_{n\beta}^{(z)}$
are the on-site energies}. \added{As in Eq. \ref{eq:epsilon_V_rho},
symmetry restricts the hopping and on-site terms to a
minimal set,}
\begin{equation}
\begin{aligned}
    & \added{h^{(z)}_{n\beta} =
    h^{(z)}_{nr_\beta}(\Gamma_\beta),}
    \\
    & \added{e^{(z)}_{n\beta} =
    e^{(z)}_{nr_\beta}(\Gamma_\beta).}
\end{aligned}
\end{equation}
\par 
Once the bands are
discretized, we \deleted{use the coupling
coefficients $h_n^z$ to} write a series of Hamiltonians 
\begin{align}
    &H_N = 
    \replaced{
        \Lambda^{(N-1)/2}
    }{
        \Lambda^{-N/2}
    }
    \replaced{
        \left[
            \sum_{n=0}^{N-1}
            h_{n\beta}^{(z)}
            (c^\dagger_{n,\beta}c_{n+1,\beta}+\text{h.c.})
            +\sum_{n=0}^N 
            e_{n\beta}^{(z)}    
            c^\dagger_{n\beta} c_{n\beta}
            +H_0
        \right],
    }{
        \left[
            \sum_{n=1}^{N}\sum_{\beta} \xi_{n}^z
            (f^\dagger_{\beta,n}f_{\beta,n-1}+\text{h.c.})
            +H_0
        \right]
    }
\end{align}
where $H_0=H_\text{occ}+H_\C+H_\text{hyb}$. This
sequence of Hamiltonians $H_N$ is iteratively constructed,
diagonalized and truncated following the symmetry-adapted
procedure described in Ref. \cite{moca2012} (see Appendix
A). Using symmetry we are able to effectively reduce the
Hilbert space by directly working with multiplets with
quantum numbers $(\Gamma_w,r_w)$, each representing a
number of states equal to $\dim (I_w) (2S_w+1)$, where $\dim
(I_w)$ is the dimension of the $I_w$ orbital irrep of the
multiplet. Moreover, the block-diagonal structure of the
Hamiltonian in the indices $\Gamma_w$ allows for an
efficient block-wise diagonalization.
\par
The code is optimized so that the overhead from the
implementation of symmetry in the block-diagonalization
procedure is minimal. By efficiently managing the reduced
matrices and using pre-computed Clebsch-Gordan factors (see
Appendix A), we are able to reduce the time spent outside of
the matrix diagonalization function to less than half of the
total computation time for most calculations. Importantly,
the relative weight of the diagonalization grows with the
size of the Hilbert space, so in demanding calculations we
obtain all the advantages of block-diagonalization at a very
little computational cost.

\section{\added{Applications}\label{Applications}}
\added{In this section we apply the \texttt{PointGroupNRG}
code to two impurity models. First we use it to solve a
two-impurity model with RKKY interaction
\cite{ruderman1954,kasuya1956,yosida1957}, which serves to
ilustrate the symmetry-adapted approach with a simple point
group and it also incorporates a non-constant DOS. Then we 
focus on a two-orbital, two-channel $E_g$ model with cubic
symmetry, which we compare with an approximate Dworin-Narath
model with continuous $SU(2)_\O$ orbital symmetry.}

\subsection{\added[comment={whole subsection
added}]{Two-impurity model with RKKY
interaction}}\label{RKKY}
The model consists of two identical $s$-shell Anderson
impurities with a separation vector $\mathbf R$ placed in flat band
of electrons with linear dispersion relation $\epsilon =
v_F(k-k_F)$ measured from the Fermi level with $v_F=D/k_F$.
The system is symmetric under inversion with respect to the
middle point between the impurities, so we work with the
$C_i$ point group. The symmetry-adapted orbitals are even
($+$) and odd ($-$) combinations of the impurity $s$
orbitals
$\psi_1(\mathbf r)=\psi(\mathbf r+\mathbf R/2)$
and 
$\psi_2(\mathbf r)=\psi(\mathbf r-\mathbf R/2)$,
\begin{equation}
    \psi_\pm(\mathbf r) = 
    \frac{1}{\sqrt 2}[\psi_1(\mathbf r)\pm\psi_2(\mathbf r)],
\label{eq:even_odd_orbitals}
\end{equation}
which belong to the irreducible representations $A_g$ (even)
and $A_u$ (odd) of $C_i$, respectively. It is worth noting
that other group choices featuring one-dimensional
irreducible representations $A$ and $B$ equivalent to even
and odd, \textit{i.e.}, fulfilling $A\otimes A=A$, $A\otimes
B=B$ and $B\otimes B=A$, are also valid in this case. In
general, what determines the application of symmetry in the
model is not the group and its group elements (symmetry
operations), but the irreducible representations appearing
in the system and their products. 
\par
Using the orbitals $\psi_\pm$, \texttt{PointGroupNRG} generates the
multiplet states, which are shown in Table \ref{tab:1-2-states_RKKY}
for up to $N=2$ particles. This information allows us to 
define the symmetry-adapted parameters of the Hamiltonian.
Starting with the impurity term $H_\text{imp}$, we assign equal
occupation energies to the orbitals, 
\begin{equation}
    \epsilon_1(\Gamma_1)=\epsilon_1(\Gamma_2)=\epsilon_\pm=\epsilon. 
\end{equation}
To determine the Coulomb parameters, we assume that the
only non-vanishing Coulomb integral (Eq.
\ref{eq:CoulombIntegral}) is the on-site repulsion
$U:=U_{1111}=U_{2222}$ in each
of the impurities. 
Then, taking into account
the orbital part of the two-particle multiplets and Eq.
\ref{eq:even_odd_orbitals}, it can be shown, by inverting
Eq. \ref{Utransformation}, that the
symmetry-adapted parameters are
\begin{equation}
\begin{aligned}
    \mathbf U(\Gamma_3) = 
    \frac{U}{2}
    \begin{pmatrix}
        1 & 1 \\
        1 & 1
    \end{pmatrix},
    \;
    U_{11}(\Gamma_4) = U,
    \;
    U_{11}(\Gamma_5) = 0.
\end{aligned}
\end{equation}
\begin{table}[!ht]
\centering
\begin{tabular}{ m{3cm}  m{1cm} c c }
\hline
$\Gamma=(N,I,S)$ & $r$ & Orbital part & Spin part   \\
\hline\hline 
$\Gamma_1=(1,A_g,\tfrac{1}{2})$ & $1$ & $\psi_+$ &
$\uparrow$, $\downarrow$ \\
$\Gamma_2=(1,A_u,\tfrac{1}{2})$ & $1$ & $\psi_-$ &
$\uparrow$, $\downarrow$ \\
$\Gamma_3=(2,A_g,0)$ & $1$ & $\psi_-\psi_-$ &
$\frac{1}{\sqrt 2}(\uparrow\downarrow-\downarrow\uparrow)$\\
                     & $2$ & $\psi_+\psi_+$ &
$\frac{1}{\sqrt 2}(\uparrow\downarrow-\downarrow\uparrow)$ \\
$\Gamma_4=(2,A_u,0)$ & $1$ & $\frac{1}{\sqrt 2}(\psi_+\psi_-
+ \psi_-\psi_+)$  &
$\frac{1}{\sqrt 2}(\uparrow\downarrow-\downarrow\uparrow)$ \\
$\Gamma_5=(2,A_u,1)$ & $1$ & $\frac{1}{\sqrt 2}(\psi_+\psi_-
- \psi_-\psi_+)$ &
$\uparrow\uparrow$, $\frac{1}{\sqrt 2}(\uparrow\downarrow
+\downarrow\uparrow)$, $\downarrow\downarrow$\\
\hline
\end{tabular}
\caption{Orbital and spin parts of the one-particle and
two-particle multiplet states of the two impurities.
Following the procedure described in Section
\ref{multiplets}, two-particle orbital states are obtained by combining 
$\psi_+\psi_+$, $\psi_+\psi_-$, $\psi_-\psi_+$ and
$\psi_-\psi_-$ so that the resulting state (i) has a
definite orbital symmetry $A_{g}$ or $A_u$, and 
(ii) is symmetrized with respect to particle
permutations. Permutation-symmetric (-antisymmetric) orbital
states are then combined with permutation-antisymmetric
(-symmetric) spin states to obtain the full permutation-antisymmetric
multiplet states.} 
\label{tab:1-2-states_RKKY}
\end{table}
\par
The hybridizations between the even and odd orbitals and the 
electron bath are given by \cite{sakai1990}
\begin{equation}
    \Delta_\pm (\epsilon) := 
    \pi \rho_\pm(\epsilon) V_\pm^2 =
    \Delta_0 
    \left(
        1\pm\frac{\sin k(\epsilon)R}{k(\epsilon)R}
    \right),
\end{equation}
where $\Delta_0$ is the hybridization at $kR\rightarrow\infty$, 
$k(\epsilon)=\epsilon/v_F+k_F$
is given by the dispersion relation, and
$\rho_\pm(\epsilon)$ and $V_\pm$ are defined by absorbing
the energy-dependence of $\Gamma_\pm(\epsilon)$ into
$\rho_\pm(\epsilon)$ and requiring that
\begin{equation}
    \int_{-D}^D d\epsilon\rho_\pm(\epsilon) = 1.
\end{equation}
Thus, $\rho_\pm(\epsilon)$ and $V_\pm$ can be regarded as
the effective density of states and hybridization amplitude
for the corresponding channel, respectively. Following their
definition, these are given by
\begin{equation}
\begin{aligned}
    &\rho_\pm (\epsilon) = 
    \frac{
        1\pm\frac{\sin k(\epsilon)R}{k(\epsilon)R}
    }{
        \int_{-D}^D d\epsilon 
        (1\pm\frac{\sin k(\epsilon)R}{k(\epsilon)R})
    },
    \\
    & V_\pm = 
    \left[
        \frac{\Delta_0}{\pi}     
        \int_{-D}^D d\epsilon 
        \left(
            1\pm\frac{\sin k(\epsilon)R}{k(\epsilon)R}
        \right)
    \right]^\frac{1}{2}.
\end{aligned}
\end{equation}
The symmetry-adapted expressions for these parameters are
\begin{equation}
\begin{aligned}
    \rho_1(\Gamma_1;\epsilon) = \rho_+(\epsilon),
    \;
    \rho_1(\Gamma_2;\epsilon) = \rho_-(\epsilon),
    \;
    V_1(\Gamma_1) = V_+,
    \;
     V_1(\Gamma_2) = V_-.
\end{aligned}
\end{equation}
\par
\begin{figure}
    \centering
    \includegraphics[width=0.8\textwidth]{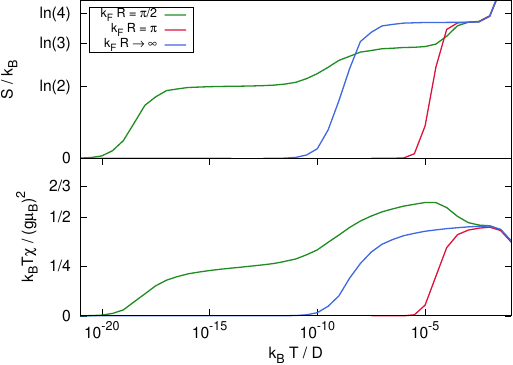}
    \caption{Entropy $S/k_B$ and magnetic susceptibility
        $k_B T \chi / (g\mu_B)^2$ curves as a
        function of temperature for the two-impurity system 
        with RKKY interaction, where $k_B$ is the
        Boltzmann factor, $g=2$ is the Landé factor, and
        $\mu_B$ is Bohr's magneton. The calculations are
        performed with fixed
        $\epsilon=-0.1$, $U=0.2$ and
        $\Delta_0=0.005$, all in units of the half-bandwidth
        $D$, and varying values
        of $k_F R=\frac{\pi}{2}$, $\pi$ and $\infty$. The values marked in
        the $y$ axis in the top (bottom) graph correspond to
        fixed points of the Hamiltonian: $\ln(4)$ and $1/2$
        correspond to the double local moment regime where 
        the impurities behave as independent
        spin-$\frac{1}{2}$ local moments, $\ln(3)$ and $3/2$
        correspond to the RKKY fixed point where the
        impurity spins are ferromagnetically coupled, and
        $\ln(2)$ and $1/4$ coorespond to the partially
        screened screened fixed point where the RKKY-coupled
        spin-$1$ is partially screened by the even channel
        (note that $\Delta_+>\Delta_-$).}
    \label{fig:thermo_rkky}
\end{figure}

 
To show the competition between the RKKY interaction and the
Kondo effect, we have computed the entropy $S$ and magnetic
susceptibility $\chi$ as a function of temperature for
various inter-impurity distances, which result in different
values of $k_F R$. For all the calculations we used a
discretization parameter $\Lambda=10$ and a cutoff of 1300
multiplets. The results are given in Fig.
\ref{fig:thermo_rkky}. For $k_F R=\frac{\pi}{2}$, the
band-induced RKKY interaction $I^{[\pi/2]}_\text{RKKY}$ is
ferromagnetic and larger than the Kondo energy scale,
$I^{[\pi/2]}_\text{RKKY}\gg k_B T^{[\pi/2]}_K$, which
results in the development of a ferromagnetically coupled
ground state at $k_B T\approx 10^{-5}D$. Since
$\Delta_+>\Delta_-$, this RKKY spin-$1$
local moment is partially screened first by the even channel
at $k_B T\approx 10^{-10}D$, and then completely screened by
the odd channel at $k_B T\approx 10^{-20}D$. For $k_F
R\rightarrow\infty$, the hybridizations are constant and equal,
$\Delta_\pm=\Delta_0$, so there is no RKKY interaction and
the spin-$\frac{1}{2}$ impurity local moments are separately
screened at an energy scale $k_B T_K^{[\infty]}\approx
10^{-8}D$. For $k_F R=\pi$, the RKKY interaction is
antiferromagnetic and larger in magnitude than the Kondo
scale, $-I^{[\pi]}_\text{RKKY}\gg k_B T^{[\pi]}_K$, which
results in the formation of an
antiferromagnetic ground state at an energy scale
$I^{[\pi]}_\text{RKKY}\approx 10^{-5}D$. The results obtained
here with the Anderson model agree with
the perturbative analysis in Ref. \cite{jayaprakash1981} and
with the NRG results obtained in Ref. \cite{silva1996} using
a Kondo model for the same system.

\deleted{Application to the two-orbital $E_g$ model}

\subsection{\added{Two-orbital $E_g$ model}} \label{Eg}
\deleted{As an example of the application of
    \texttt{PointGroupNRG}, we
investigate an impurity system with two $E_g$ orbitals.
This}
\added{We now investigate a model consisting of two orbitals
belonging to the irreducible representation $E_g$ of the
cubic point group $O_h$, each one connected to a channel of
the same symmetry. This model}
can be regarded as a transition metal impurity with the
outer $d$ shell degeneracy split by a crystal field with cubic
symmetry $O_h$ into a three-fold degenerate
$T_{2g}$ upper level and a two-fold degenerate $E_g$ lower
level, where the energy difference between the two of them is
sufficiently large to discard the $T_{2g}$ subspace
altogether.
\par
For the orbital part, we use the basis functions and Clebsch-Gordan
coefficients found in the tables by Altmann and
Herzig \cite{Altmann}. The orbital basis functions are
\begin{equation}
\begin{aligned}
\label{eq:basis}
    &\psi_1(\mathbf r)
    =
    \frac{1}{\sqrt 2}
    [\psi_a(\mathbf r) - i\psi_b(\mathbf r)]
    \\
    &\psi_2(\mathbf r)
    = 
    \frac{1}{\sqrt 2}
    [\psi_a(\mathbf r) + i\psi_b(\mathbf r)]
\end{aligned}
\end{equation}
where $\psi_a(\mathbf r)=R(r)Y^0_2(\theta,\varphi)$ and
$\psi_b(\mathbf r) = R(r)[Y^2_2(\theta,\varphi) +
Y^{-2}_2(\theta,\varphi)] /\sqrt 2$ are real harmonics with
a radial term $R(r)$.
\par
\begin{table}[!ht]
\centering
\begin{tabular}{ m{3cm}  m{1cm}   m{1cm}  c }
\hline
$\Gamma\added{=(N,I,S)}$ & $T$ & $i$ & Orbital part   \\
\hline\hline 
$\Gamma_1=(1,E_g,\tfrac{1}{2})$ & $\frac{1}{2}$ & 1 &
    $\psi_1$  \\
&  & 2 & $\psi_2$ \\
$\Gamma_2=(2,E_g,0)$ & 1 & 1 & $\psi_2\psi_2$  \\
& 1 & 2 & $\psi_1\psi_1$  \\
$\Gamma_3=(2,A_{1g},0)$  & 1 & 1 & $\frac{1}{\sqrt 2}(\psi_1\psi_2+\psi_2\psi_1)$  \\
$\Gamma_4=(2,A_{2g},1)$  & 0 & 1 & $\frac{1}{\sqrt
2}(\psi_1\psi_2-\psi_2\psi_1)$ \\
\hline
\end{tabular}
\caption{Orbital structure of the one-particle and two-particle multiplet states 
of the $E_g$ impurity. Column $\Gamma$ specifies the irrep,
$T$ is the orbital isospin, and $i$ is the orbital partner
label.} 
\label{tab:1-2-states}
\end{table}
\par 
We now have to find the independent parameters of the
Hamiltonian. In this case, the only one-body irrep is
$\Gamma_1=(1,E_g,\tfrac{1}{2})$, and the two-body
irreducible representations
are $\Gamma_2=(2,E_g,0)$, $\Gamma_3=(2,A_{1g},0)$ and 
$\Gamma_4=(2,A_{2g},1)$; we use these irreducible representations 
to label the
corresponding multiplets. \added{As in Section \ref{RKKY},
we use the orbital part of multiplet states provided by 
\texttt{PointGroupNRG} and given in Table
\ref{tab:1-2-states} to obtain the independent parameters}
\deleted{\texttt{PointGroupNRG} provides the
exact form of the many-particle multiplet states in a first
output of the calculation. Taking spin-orbital symmetry into
account, the independent parameters are}
\begin{equation}
\begin{aligned}
    \epsilon:=\epsilon_1(\Gamma_1),\;\;
    V:=V_{11}(\Gamma_1),\;\;
    U_2:=U_{11}(\Gamma_2),\;\;
    U_3:=U_{11}(\Gamma_3),\;\;
    U_4:=U_{11}(\Gamma_4).
\end{aligned}
\end{equation}
\par
We can relate our independent Coulomb parameters to the
usual Coulomb integrals obtained in the basis of real
orbitals $\psi_a(\r)$ and $\psi_b(\r)$. The only
non-vanishing terms are the intra-orbital repulsion
$U:=U_{aaaa}$, the inter-orbital repulsion $U':=U_{abba}$,
the exchange $J:=U_{abab}$, and the pair-hopping
$J':=U_{aabb}$ (note that $a$ and $b$ can be interchanged).
Using Eq. \ref{eq:basis} and the expressions of the
multiplet states in Table \ref{tab:1-2-states}, we obtain
the general expressions shown in Table \ref{tab:Coulomb}. If
(i) we impose the physical condition $J=J'$ that results
from the permutation symmetry in the basis of real orbitals
and (ii) we use the selection rule that forbids transitions
between different partners of irrep $\Gamma_2$, we obtain an
additional set of equivalent restrictions, namely, $2U_2=U_3+U_4$
and $U'=U-2J$. For an $E_g$ subspace, these restrictions are
the same as for Coulomb interactions with cubic and full
rotational symmetries \cite{castellani1978,bunemann2017}.
This leaves us with two independent Coulomb parameters for
the $O_h$-symmetric system, as shown in Table \ref{tab:Coulomb}.
\begin{table}[!ht]
\begin{tabular}{ m{2cm} m{4cm} m{3cm} m{3cm} }
\hline 
& \text{General} & $O_h$, $J'=J$ & $SU(2)_\O$, $J'=0$ \\
      \hline\hline
    $U_2$ & $\frac{1}{2}(U+U'+J-J')$ &  $U+J$ & $U$ \\ 
     $U_3$ & $U+J'$ & $U-J$ & $U$ \\ 
     $U_4$ & $U'-J$ & $U-3J$& $U-2J$ \\
    \hline
\end{tabular}
\caption{Relation between symmetry-adapted Coulomb
parameters and the usual Coulomb integrals for the general
case, \textit{i.e.}, before applying orbital symmetries and
the restrictions on $J'$, for the $O_h$-symmetric model and
for the $SU(2)_\O$-symmetric model.}
\label{tab:Coulomb}
\end{table}
\par
As a first step, let us consider the application of
continuous symmetries to this system. For a full orbital
symmetry $SU(2)_\O$ we need to set $U_2=U_3$ because the
orbital irreps $\Gamma_1$ and $\Gamma_2$ irreps have the
same orbital isospin $T=1$ (see Table \ref{tab:1-2-states}).
This condition is fulfilled in our $O_h$-symmetric model by
setting $J=0$, which removes the Hund physics from the
system. An alternative is to ignore the pair-hopping terms
by setting $J'=0$ and $J\neq J'$; this choice gives us the
Dworin-Narath Hamiltonian \cite{dworin1970,georges2013}
used in a previous work for a similar two-orbital $E$
model \cite{barral2017}. It has the advantage of exhibiting
Hund physics while preserving $SU(2)_\O$ symmetry, but it
must be emphasized that it is only an approximation, because
by choosing $J'=0$ we break the permutation symmetry of the
Coulomb integrals in the basis of real orbitals. 
\par
\begin{figure}[!ht]
    \label{fig:mag_Eg}
    \centerline{\includegraphics[width=0.8\textwidth]{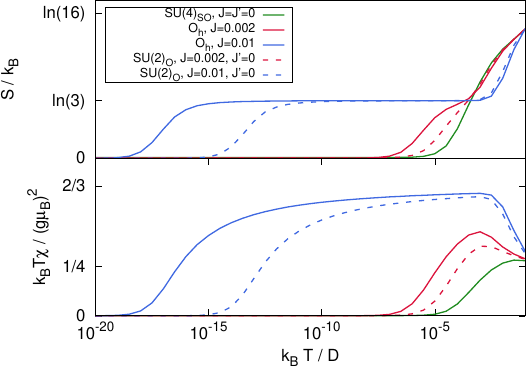}}
    \caption{Entropy and magnetic susceptibility curves as a
        function of temperature \added{for the two-orbital $E_g$
        system} for fixed
        $\epsilon=-U=-0.1$ and
        $\replaced{\Delta}{\rho \pi V^2}=0.002$ with varying values
        of $J=0$, $0.002$ and $0.01$. The values marked in
        the $y$ axis in the top (bottom) graph correspond to
        fixed points of the Hamiltonian: $\ln(16)$ and $1/4$
        correspond to the free orbital regime where all the
        impurity states are degenerate, $\ln(3)$ and $3/2$
        correspond to the fixed point where the impurity
        behaves as a local moment with spin $S=1$.}
\end{figure}

In order to examine the numerical difference between the
$O_h$-symmetric model and the $SU(2)_\O$-symmetric model, we
perform entropy $S$ and magnetic susceptibility $\chi$
calculations for both systems with varying $J$, fixed
$\epsilon=-U=-0.1$, and a constant hybridization $\Delta =
\rho \pi V^2 = 0.002$ \replaced{and DOS}{with} $\rho=1/2$,
all in units of the half-bandwidth $D$. The parameters are
chosen so as to showcase the Hund physics of the systems:
for $J>0$ the $S=1$ states belonging to irrep $\Gamma_4$
have the lowest energy; for $J=0$ both models are equivalent
and they possess $SU(4)_\SO$ spin-orbital symmetry. We set
the discretization parameter to $\Lambda=10$ and we average
over the results obtained with $z=0$ and $z=1/2$.
At each
iteration we keep a minimum of
600 multiplets, which amounts to approximately 1500 states
on average. In Fig. \ref{fig:mag_Eg} we show the calculated
impurity contributions to the entropy and magnetic
susceptibility as a function of temperature. For both
systems, we observe that by increasing $J$ the system gets
closer to the spin-$1$ local moment fixed point, where $k_B
T\chi /(g\mu_B)^2=2/3$\deleted{ with a Landé factor $g=2$}, and
$S/k_B=\ln(3)$, before finally reaching the low-temperature
strong coupling fixed point where $\chi=S=0$. This leads to a
lower Kondo temperature, as expected from previous
studies \cite{georges2013}. Comparing the curves of the two
models for the same parameter choices, we see that the
$O_h$-symmetric model has lower Kondo temperatures
than the $SU(2)_\O$-symmetric model. This is a result of the
difference in the Coulomb parameters (see Table
\ref{tab:Coulomb}), which increases the energy gap between
the $\Gamma_4$ ground multiplet and the multiplets $\Gamma_1$
and $\Gamma_5=(3,E_g,\tfrac{1}{2}$) in the $O_h$
case as compared to the $SU(2)_\O$ case; these are the
multiplets that contribute to the Kondo coupling
mechanism up to second order in perturbation theory. In
conclusion, having a larger energy gap means that the
antiferromagnetic coupling is weaker \cite{schrieffer1966}.
\par

\section{\added[comment={whole section
added}]{Benchmarks}}\label{Benchmarks}
In this section we provide two types of performance benchmarks 
for the code. First we measure the improvements in speed and 
memory efficiency obtained when using \verb|PointGroupNRG|
with orbital symmetries. Then we compare the performance of
\texttt{PointGroupNRG} versus \texttt{NRGLjubljana}
\cite{nrgljubljana}, a widely
used \cite{skolimowski2018,huang2023,zonda2015,requist2014}
open-source NRG code. All calculations are performed on a
computer with an Intel(R) Core(TM) i7-10750H processor and
16GB of RAM. For the benchmark, the
\texttt{PointGroupNRG} code has been compiled and saved into
a sysimage using \texttt{PackageCompiler.jl}, which removes
the latency due to package loading and function compilation
\cite{packagecompiler}. Scripts to perform this
precompilation are provided with the code. As performance
indicators, we use the elapsed time and maximum resident set
size (max. RSS) for each serial run with an increasing
number of multiplet cutoffs. The max. RSS is a measure of
the peak memory usage of the process. For all calculations,
the number of NRG iterations is 42 and the discretization
parameter is $\Lambda=3$.

\subsection{Performance improvement with
symmetry}\label{benchmarks_symmetry}
\begin{figure}[ht!]
    \centering
    \includegraphics[width=1.0\textwidth]{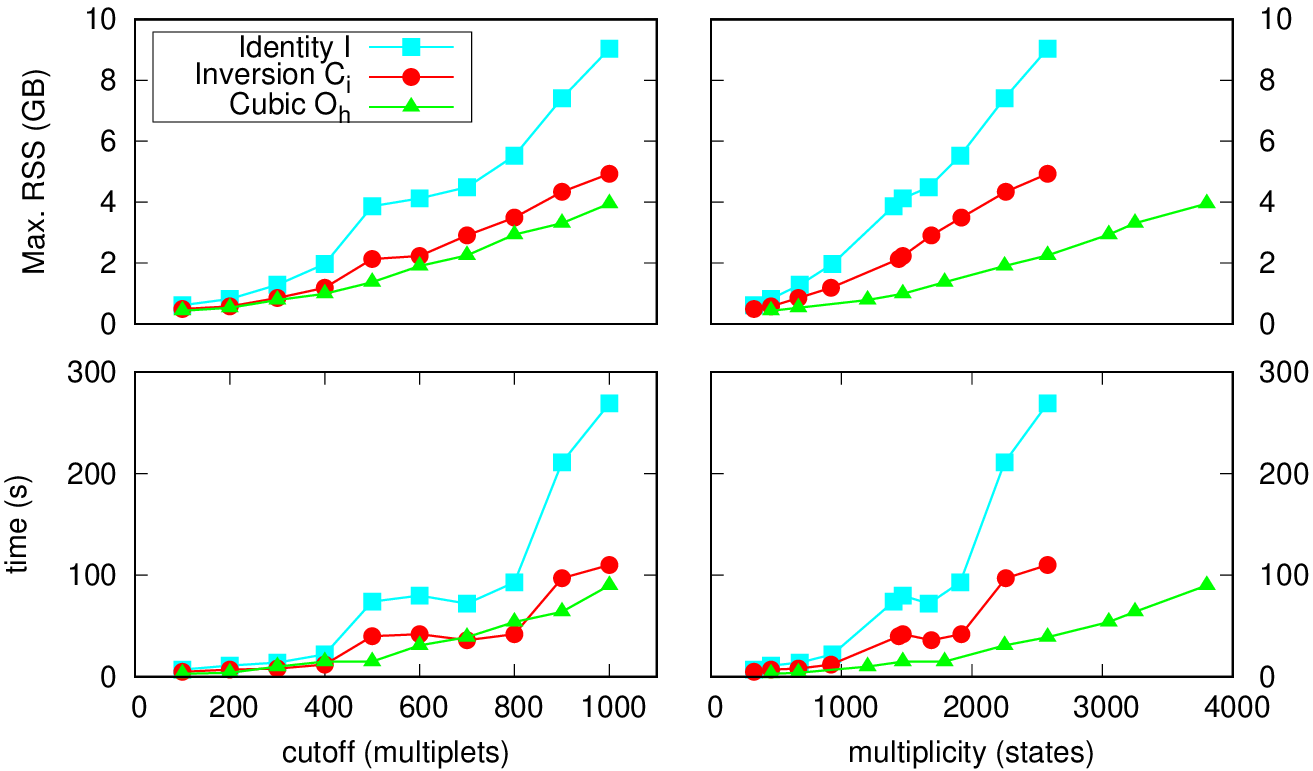}
    \caption{Time (bottom panels) and maximum resident set size
    (top panels) for the two-impurity, two-channel Anderson Hamiltonian
    for various multiplet cutoffs. The results are shown as
    a function of the cutoff (left panels) and as a function
    of the average number of states represented at the last two steps 
    for the corresponding cutoff (right panels).}
    \label{fig:benchmark_symmetry}
\end{figure}

To quantify the effect of considering symmetries by the methods 
described in this work, we measure the total computation time 
and the peak
memory usage in calculations for increasingly larger orbital
point groups: the idenitity point group $I$ 
(no orbital symmetry), the inversion group $C_i$ with one
$A_{1g}$ orbital and one $A_{1u}$ orbital (as in Sec.
\ref{RKKY}), and the
cubic group $O_h$ with two $E_g$ orbitals (see Sec.
\ref{Eg}). We apply them to the same model
consisting of two orbitals, with a corresponding impurity
Hamiltonian $H_\text{imp}=H_\text{occ}+H_\C=0$, and two channels, each coupled
to one orbital with an equal hybridization $\Delta=0.1D$. The results 
are shown in Fig.
\ref{fig:benchmark_symmetry} for various multiplet cutoffs. \par
For a given multiplet cutoff, it can be seen that the peak
memory usage is reduced as the exploited symmetry becomes
larger. Regarding speed, an improvement results from using 
$C_i$ or $O_h$ over $I$, but no appreciable difference
results from choosing $O_h$ over $C_i$. Overall, these
results meet two expected behaviors: (i) the diagonalization is
faster because, by applying a larger orbital symmetries, the 
(already block-diagonal) reduced matrix $H_N$ for each NRG 
iteration $N$ is split into a larger number of blocks,
each block corresponding to an irreducible representation $\Gamma_u$
of the point group $P$; (ii) less memory is needed when
updating the Hamiltonian at each step, as the newly added
conduction shell subspace is represented by fewer
multiplets.
\par
If we compare the results as a function of the multiplicity
(Fig. \ref{fig:benchmark_symmetry} right column),
\textit{i.e.} the number of
states represented by the multiplets kept at each iteration,
the advantage of exploiting $O_h$ over $C_i$ becomes apparent. 
Compared to the results obtained when
comparing runs with equal multiplet cutoffs (Fig.
\ref{fig:benchmark_symmetry} left column), the time and
memory gains
are explained by the fact that the number of
states represented by each of the multiplets grows by a
factor equal to the
dimension $\text{dim}(\Gamma_u)$ of the irreducible
representation $\Gamma_u$ to which the multiplet belongs.
Therefore, the same physical cutoff can be
treated in a more compact and efficient manner when the
irreducible representations have a larger dimension. In this
case, with $O_h$ we have $E_g$ with $\text{dim}(E_g)=2$,
which explains the difference with respect to the $I$ and
$C_i$ cases, where all the orbital irreducible
representations are one-dimensional.

\subsection{Comparison with another
code}\label{benchmarks_comparison}
As a performance standard we choose
\texttt{NRGLjubljana}\cite{nrgljubljana,zitko2009}. Part of
\texttt{NRGLjubljana} is written in Mathematica, which is
used for the preparation of the basis, the operators and the
initial Hamiltonian, and part in C++, where the numerically
intensive sections of the procedure are performed.
\par
\begin{figure}[ht!]
    \centering
    \includegraphics[width=1.0\textwidth]{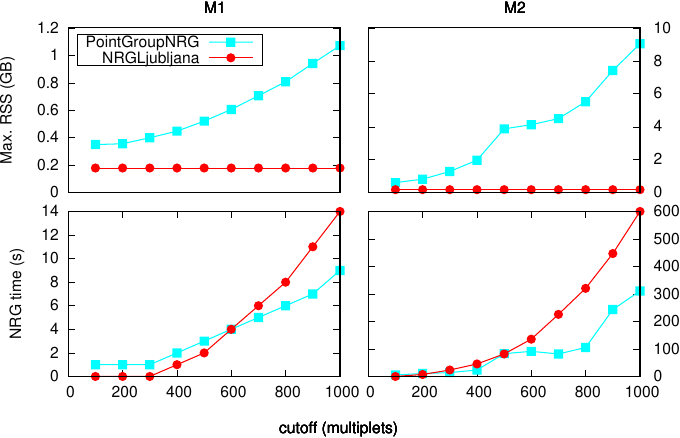}
    \caption{NRG calculation time (bottom panels) and
    maximum resident set size (top panels) for the
    one-orbital, one-channel model (left panels) and the
    two-orbital, two-channel model (right panels) for
    various multiplet cutoffs.}
    \label{fig:pgnrg_vs_nrgl}
\end{figure}

 
For the
comparison of \texttt{PointGroupNRG} with
\texttt{NRGLjubljana}, we have performed calculations for a
single-orbital, single-channel model (M1) and a model with two
orbitals and channels (M2). The model has $H_\text{imp}=0$ and a
hybridization $\Delta=0.1D$ for every channel. With both
codes we use the symmetry $U(1)_\C\otimes SU(2)_\S$, which
is achieved in \verb|PointGroupNRG| by setting the orbital
symmetry to be the identity group and in \verb|NRGLjubljana| by
choosing the tag \texttt{symtype=QS}. The results are shown in Fig.
\ref{fig:pgnrg_vs_nrgl}.
\par
Peak memory (max. RSS) results show that
\texttt{PointGroupNRG} requires significantly more memory
than \texttt{NRGLjubljana}. 
The larger max. RSS is partially attributed to the memory management
of the Julia language \cite{garbagecollector}. 
The computation time is of the same
order of magnitude overall for both codes with the chosen multiplet
cutoffs, with \texttt{PointGroupNRG} being faster for
large cutoffs. With the addition of symmetry, the speed is
further improved and the memory demands are diminished, as
shown in Fig. \ref{fig:benchmark_symmetry}.

\section{Conclusions}
We have presented our Julia code, \texttt{PointGroupNRG}, which
provides a simple and flexible framework for constructing
and solving magnetic impurity models with finite point group
symmetries. By exploiting the existing orbital symmetries, 
it allows efficient and precise
NRG calculations without the need to work with approximate
models with continuous symmetry. We have obtained results 
using Anderson impurity Hamiltonians with up to two
channels. Models of this kind can be used for transition metal
impurities in crystalline environments. This includes, for
example,
adatoms on metallic surfaces \cite{baruselli2015}, impurities
in nanotubes \cite{baruselli2013}, and impurities in metallic
chains \cite{dinapoli2015,barral2017,blesio2019}
\added{, and it can also be applied to several quantum dot
systems\cite{tanaka2012,karki2018,chen2021}. We have showcased the \texttt{PointGroupNRG} code by
    applying it to a parity-symmetric two-impurity model
    with energy-dependent DOS exbiting RKKY interaction and
    to a two-orbital $E_g$ model with cubic $O_h$ symmetry.}
\deleted{Parity-symmetric two-impurity systems can also be solved by
this general procedure using inversion or reflection
symmetry groups, \textit{e.g.}, the two-impurity system with
RKKY interaction \cite{silva1996} and several quantum dot
systems} 
\par
The \texttt{PointGroupNRG} code can be also be extended to
cover different models. One such extension would be to
include spin-orbit coupling, which can be achieved by using
a double group instead of the orbital point group. This
would make it possible to apply the code to $f$-shell
impurity systems.
\par
\added{As shown by the performance benchmarks, the code
allows for fast calculations for arbitrary models with
any point group symmetry. We have also demonstrated the
advantage of using orbital symmetries, which results in
faster and more memory-efficient calculations.}

\section*{Acknowledgements}
We acknowledge grants No. IT-1527-22, funded by the
Department of Education, Universities and Research of the
Basque Government, No. PID2019-103910GB-I00 funded by
MCIN/AEI, 10.13039/501100011033/, and No. PRE2020-092046
funded by the Ministry of Education of Spain.

\appendix
\section*{Appendix A.}\label{App.A}
The core of the symmetry-adapted approach in terms of
iterating through NRG steps is the reduced matrix
representation, which relies on the Wigner-Eckart theorem.
The latter states that for an operator $\mathcal O_{a}$ that
transforms as the $\gamma_a$ partner of irrep $\Gamma_a$,
its matrix elements in a basis of symmetry-adapted 
states $\ket{u}$, $\ket{v}$ fulfill 
\begin{equation}
    \bra{u} \mathcal O_a \ket{v}
    =
    \bra{m_u}| \mathcal O_{m_a} |\ket{m_v}
    (\Gamma_a,\gamma_a;\Gamma_v,\gamma_v|
    \Gamma_u,\gamma_u,r_u)^*,
\end{equation}
where $\bra{m_u}| \mathcal O_{m_a} |\ket{m_v}$ is a reduced
matrix element that depends only on the multiplet quantum
numbers $m_u=(\Gamma_u,r_u)$ and $m_v=(\Gamma_v,r_v)$. 
\par 
When diagonalizing the Hamiltonian at each step, the reduced
representation provides a two-fold advantage. On the one
hand, since the Hamiltonians $H_N$ are fully symmetric with
respect to $ G$, the matrix elements are
block-diagonal in the irreps,
$\langle\replaced{m_u}{u}||H_N|| 
\replaced{m_v}{v}\rangle\propto
\delta_{\Gamma_u,\Gamma_v}$. On the other hand, each of the
blocks is smaller by a factor of $\text{dim}(\Gamma_u)$
than the blocks that would result from the full matrix. This
makes the calculations faster and, since the degeneracies
are treated exactly, also more precise. Our
\verb|PointGroupNRG| code implements the general expressions
given in Ref. \cite{moca2012} adapted to the case of a
finite symmetry group $G$. Below we give the main equations.
\par 
In order to update the Hamiltonian, we first decompose the
reduced matrix elements as 
\begin{equation}
    \bra{m_u}|H_N|\ket{m_v}_{(\replaced{N}{n})}
    = 
    \delta_{m_u,m_v}
    \left(
        \Lambda^{1/2} E_{m_i}^{(\replaced{N}{n})} 
        +
        \added{\bra{m_\mu}|\mathcal E_N|\ket{m_\mu}}
    \right)
    + 
    \bra{m_u}|\tau_{\replaced{N}{n}}|\ket{m_v}_{(\replaced{N}{n})}.
\end{equation}
\added{The diagonal elements are given by
$\Lambda^{1/2}E^{(\replaced{N}{n})}_{m_i}$, the rescaled
eigenenergy of the $m_i$ multiplet of step
$\replaced{N}{n}$, and the rescaled shell occupation energy
$\bra{m_\mu}|\mathcal E_N|\ket{m_\mu}$, where}
\begin{equation}
    \added{
        \mathcal E_N
        =
        \Lambda^\frac{N}{2}
        \sum_{\beta} 
        e^{(z)}_{N r_\beta}
        c^\dagger_{N \beta}c_{N \beta}
    }
\end{equation}
\added{and $m_\mu$ and $m_\nu$ are multiplets of the newly added
shell.}
\deleted{The hopping
operator $\tau_{n-1,n}$ is  defined as}
\added{The non-diagonal terms are given by the matrix elements of 
the hopping operator $\tau_{N}$, which is defined as}
\begin{equation}
    \tau_{\replaced{N}{n-1,n}} 
    = 
    \added{\Lambda^\frac{N-1}{2}}
    \sum_{\replaced{\beta}{a}} 
    \replaced{
        h^{(z)}_{N-1,r_\beta}(\Gamma_\beta)
    }{
        \xi^{(n)}_{r_a}(\Gamma_a) 
    }
    \replaced{
        c^\dagger_{N-1,\beta}c_{N \beta}
    }{
        c^\dagger_{a,n-1}c_{a,n}
    }
    + \hc.    
\end{equation} 
\par
The multiplets to be diagonalized in step $n$ are
obtained by combining shell multiplets, which contain only
electrons in the newly added shell, and block multiplets,
which combine electrons from all the previous shells. The
notation is as follows. For a $\replaced{N}{n}$-th step matrix element,
marked with a subscript \deleted{n}$\added{N}$, the multiplets $m_u$ ($m_v$)
to be diagonalized are obtained by combining block
multiplets $m_i$ ($m_v$) with shell multiplets $m_\mu$
($m_\nu$). Following this convention, the hopping matrix
element is constructed as
\begin{equation}
\begin{aligned}
    \label{eq:redmatconstruction}
    \bra{m_u}|\tau_{\replaced{N}{n-1,n}}&|\ket{m_v}_{(\replaced{N}{n})}
    = 
    \delta_{\Gamma_u,\Gamma_v} 
    (-1)^{N_\mu}
    \sum_{m_{\replaced{\beta}{a}}} 
    h_{\added{N}r_{\replaced{\beta}{a}}}^{(\replaced{z}{n})}(\Gamma_{\replaced{\beta}{a}}) 
    \bra{m_\nu}|
    c^\dagger_{\added{N}m_{\replaced{\beta}{a}}\deleted{,n}}
    |\ket{m_\mu}_{(\replaced{N}{n})}^* \\
    &\times\bra{m_{i}}|
    c^\dagger_{\added{N-1,}m_{\replaced{\beta}{a}},\deleted{n-1}}
    |\ket{m_{j}}_{(\replaced{N}{n})}
    D(\Gamma_u,\Gamma_v,\Gamma_i,\Gamma_j,
      \Gamma_\mu,\Gamma_\nu,\Gamma_a) 
    + \hc,
\end{aligned}
\end{equation}
Here $N_\mu$ is the number of particles in the $m_\mu$
multiplet and $\bra{m_\nu}|
c^\dagger_{\added{N}m_{\replaced{\beta}{a}},\deleted{n}}
|\ket{m_\mu}_{(\replaced{N}{n})}$ is a reduced matrix
element involving only shell degrees of freedom and
therefore independent of the step $\replaced{N}{n}$, so it
is computed once and used at every step. The reduced matrix
element involving the block multiplets $m_i$ and $m_j$ is
computed with the information from the previous step: if we
denote the multiplets of the diagonal basis of any given
iteration as $m_{u'}$ ($m_{v'}$), then the matrix elements
between $m_i$ and $m_j$ are just
$\bra{m_i}|c^\dagger_{\added{N-1,}m_{\replaced{\beta}{a}},
\deleted{n-1}} 
|\ket{m_j}_{(\replaced{N}{n})}
=\bra{m_{u'}}|
\replaced{c}{f}^\dagger_{\added{N-1,}m_{\replaced{\beta}{a}},
\deleted{n-1}}
|\ket{m_{v'}}_{\replaced{(N-1)}{[n-1]}}$. We decompose the
block matrix elements as
\begin{equation}
\begin{aligned}
    \label{eq:blockpart}
    \bra{m_{u'}}|
    c^\dagger_{\added{N-1,}m_{\replaced{\beta}{a}}\deleted{n-1}} 
    |\ket{m_{v'}}_{\replaced{(N-1)}{[n-1]}}
    = 
    &\sum_{r_u,r_v} 
    [U^{\replaced{(N-1)}{[n-1]}}_{r_u,r_u'}(\Gamma_u)]^* 
    U^{\replaced{(N-1)}{[n-1]}}_{r_v,r_v'}(\Gamma_v) 
    \delta_{m_i,m_j}
    \\
    &\times
    \bra{m_\mu}|
    c_{\added{N-1,}m_{\replaced{\beta}{a}}}^\dagger
    |\ket{m_\nu}_{\replaced{(N-1)}{[n-1]}}
    K(\Gamma_u,\Gamma_v,\Gamma_i,\Gamma_\mu,
        \Gamma_\nu,\Gamma_a),
\end{aligned}
\end{equation}
where $U^{\replaced{(N-1)}{[n-1]}}(\Gamma_u)$ is the unitary matrix that
diagonalizes the subspace of irrep $\Gamma_u$ in step
$\replaced{N}{n}-1$.
The coefficients $D$ and $K$ in Eqs.
\ref{eq:redmatconstruction} and \ref{eq:blockpart} are
summations over Clebsch-Gordan indices,
\begin{equation}
\begin{aligned}
    &D(\Gamma_u,\Gamma_v,\Gamma_i,\Gamma_j,
    \Gamma_\mu,\Gamma_\nu,\Gamma_a) 
    =
    \sum_{\gamma_\mu,\gamma_i}
    \sum_{\gamma_\nu,\gamma_j}
    \sum_{\gamma_a}
    (\Gamma_\nu,\gamma_\nu;\Gamma_j,\gamma_j|
    \Gamma_v,\gamma_v,r_v) 
    \\
    &\times(\Gamma_\mu,\gamma_\mu;\Gamma_i,\gamma_i|
    \Gamma_u,\gamma_u,r_u)^*
    (\Gamma_a,\gamma_a;\Gamma_\mu,\gamma_\mu|
    \Gamma_\nu,\gamma_\nu,r_\nu)
    (\Gamma_a,\gamma_a;\Gamma_\mu,\gamma_\mu|
    \Gamma_\nu,\gamma_\nu,r_\nu)^*,
\end{aligned}
\end{equation}
\begin{equation}
\begin{aligned}
    &K(\Gamma_{u},\Gamma_{v},\Gamma_{i},
      \Gamma_{\mu},\Gamma_{\nu},\Gamma_a)
      =
      \frac{1}{\text{dim}(\Gamma_u)}
      \sum_{\gamma_a}
      \sum_{\gamma_u,\gamma_v}
      \sum_{\gamma_\mu,\gamma_\nu}
      \sum_{\gamma_i}
      (\Gamma_{\nu},\gamma_{\nu};\Gamma_{i},\gamma_{i}|
       \Gamma_{v},\gamma_{v},r_{v}) \\
     &\times
      (\Gamma_{\mu},\gamma_{\mu};\Gamma_{i},\gamma_{i}|
       \Gamma_{u},\gamma_{u},r_{u})^* 
      (\Gamma_a,\gamma_a;\Gamma_{v},\gamma_{v}|
       \Gamma_{u},\gamma_{u},r_{u})
      (\Gamma_a,\gamma_a;\Gamma_{\nu},\gamma_{\nu}|
       \Gamma_{\mu},\gamma_{\mu},r_{\mu})^*.
\end{aligned}
\end{equation}
\par 
In order to compute the spectral function, we also need
$\bra{m_{u'}}|f^\dagger_{m_\alpha}|\ket{m_{v'}}_{(\replaced{N}{n})}$, where
$f^\dagger_{m_\alpha}$ creates an electron at the
one-electron impurity state $\ket{\alpha}$ and $m_{u'}$ are
$m_{v'}$ are the multiplets obtained after the
diagonalization in step $n$. The expression
in this case is 
\begin{equation}
\begin{aligned}
    \bra{m_{u'}}|
    f_{m_\alpha}^\dagger|
    \ket{m_{v'}}_{(\replaced{N}{n})}
    =&
    (-1)^{N_\mu}
    \sum_{r_u,r_v}
    \left[U^{(\replaced{N}{n})}_{r_u,r_{u'}}(\Gamma_u)\right]^*
    U^{(\replaced{N}{n})}_{r_v,r_{v'}}(\Gamma_v)
    \delta_{m_\mu,m_\nu}
    \\
    \times &
    \bra{m_i}|f^\dagger_{m_\alpha}|\ket{m_j}_{(\replaced{N}{n})}
    F(\Gamma_u,\Gamma_v,\Gamma_i,\Gamma_j,
      \Gamma_\mu,\Gamma_\alpha),
\end{aligned}
\end{equation}
where the block matrix element is obtained in the previous
step,
$\bra{m_i}|f^\dagger_{m_\alpha}|\ket{m_j}_{(\replaced{N}{n})} = 
\bra{m_{u'}}|f^\dagger_{m_\alpha}|\ket{m_{v'}}_{\replaced{(N-1)}{[n-1]}}$,
and the coefficient $F$ is a sum over Clebsch-Gordan
coefficients,
\begin{equation}
\begin{aligned}
    F(&\Gamma_u,\Gamma_v,\Gamma_i,\Gamma_j,
      \Gamma_\mu,\Gamma_\alpha)
    =
    \frac{1}{\text{dim}(\Gamma_{ u})}
    \sum_{\gamma_u,\gamma_v}
    \sum_{\gamma_i,\gamma_j}
    \sum_{\gamma_\mu}
    \sum_{\gamma_\alpha} 
    (\Gamma_\mu,\gamma_\mu;\Gamma_j,\gamma_j|
     \Gamma_v,\gamma_v,r_v)
    \\
    \times &
    (\Gamma_\mu,\gamma_\mu;\Gamma_i,\gamma_i|
     \Gamma_u,\gamma_u,r_u)^* 
    (\Gamma_\alpha,\gamma_\alpha;\Gamma_v,\gamma_v|
     \Gamma_u,\gamma_u,r_u)
    (\Gamma_\alpha,\gamma_\alpha;\Gamma_j,\gamma_j|
     \Gamma_i,\gamma_i,r_i)^*.
\end{aligned}
\end{equation}
Our code precomputes the the coefficients $D$, $K$ and $F$
for a set of irrep combinations chosen so that they cover
all the possible inputs of a given NRG run. In this way, all
the information about the structure of each irrep can be
accessed at every step with very little computational
cost.



\bibliographystyle{elsarticle-num-names} 
\bibliography{artikuloa_elsevier_review1}


%
%
%
\end{document}